# An Evaluation of Impacts in "Nanoscience & nanotechnology:" Steps towards standards for citation analysis

Loet Leydesdorff,
Amsterdam School of Communications Research (ASCoR), University of Amsterdam,
Kloveniersburgwal 48, 1012 CX Amsterdam, The Netherlands;
loet@leydesdorff.net; http://www.leydesdorff.net.

**Abstract**

One is inclined to conceptualize impact in terms of citations per publication, and thus as an average. However, citation distributions are skewed, and the average has the disadvantage that the number of publications is used in the denominator. Using hundred percentiles, one can integrate the normalized citation curve and develop an indicator that can be compared across document sets because percentile ranks are defined at the article level. I apply this indicator to the set of 58 journals in the ISI Subject Category of "Nanoscience & nanotechnology," and rank journals, countries, cities, and institutes using non-parametric statistics. The significance levels of results can thus be indicated. The results are first compared with the ISI-Impact Factors, but this Integrated Impact Indicator (*I3*) can be used with any set downloaded from the (*Social*) *Science Citation Index*. The software is made publicly available at the Internet. Visualization techniques are also specified for evaluation by positioning institutes on Google Map overlays.

## Introduction

In recent decades, scientometric indicators have increasingly been used for evaluative purposes both in public-policy domains and, for example, in hiring and promoting faculty. Policy makers and research managers need to rely on the quality of these indicators. Recent years, however, have witnessed fierce debate about the use of statistics and standards in evaluation research (Gingras & Larivière, 2011). A transition from parametric statistics (using averages) towards non-parametric statistics (using percentiles) provides advantages, but implies a different conceptualization of "impact."

Non-parametric statistics enable us to honour both productivity and quality, whereas the impact may be lower in the case of averaging for the sole reason of a higher productivity. These statistics share this appreciation of both productivity and citation rates with the *h*-index (Hirsch, 2005), but they differ from the *h*-index in that a range of tests for the significance of the impact (above or below expectation) becomes available. Less-cited papers can thus be appreciated proportionally, while the *h*-index uses the *h*-value as a threshold for the cutoff of the tails of the distributions (cf. Glänzel, 2007; Vinkler, 2010 and 2011).

Scientometric indicators first require normalization because of differences in publication and citation practices among fields of science. For example, impact factors in psychology are on average larger than in sociology by an order of magnitude (Leydesdorff, 2008). Mathematicians provide short reference lists (on the order of five references), while life-scientists often provide more than 40 references. The chances of being highly-cited and positively evaluated would thus vary among the sciences for statistical reasons (Garfield, 1979a).

Two conventions have been devised in the past for normalization: (1) normalization in terms of fields of science, and (2) comparison of the citation rate with the world average. For the delineation of fields of science, scientometricians often turn to journal classification schemes provided by the database producer—such as the ISI Subject Categories of the *(Social) Science Citation Index* (SCI and SSCI).[1] Journals, however, can be assigned to more than a single category, and this categorization is not "literary warranted"; that is, it is not updated systematically to follow the development of the scientific literature (Bensman & Leydesdorff, 2009; Chan, 1999).

The sources of error in these field delineations are difficult to control using journals as units of analysis because it is questionable whether journals themselves are mono-disciplinary units in terms of their content (Boyack & Klavans, 2011; Pudovkin & Garfield, 2002; Rafols & Leydesdorff, 2009). Furthermore, one can expect the institutional units under study—such as universities—to be organized disciplinarily and interdisciplinarily, but from a different perspective. Therefore, refinement of the journal categories cannot solve the problem of field delineation (Leydesdorff, 2006; Rafols *et al.*, in press). Perhaps, paper-based classifications such as the Medical Subject Headings (MeSH) of the Medline database would be more appropriate, but the complexity of the handling of citations in relation to more than a single database has hitherto been technically difficult (Bornmann *et al.*, 2008; 2011).

[1] Scopus of Elsevier provides similar classification schemes.

The second problem—that is, the comparison to a world average—was addressed early in the scientometric enterprise by Schubert & Braun (1986), who proposed comparing the mean observed citation rate (*MOCR*) with the corresponding mean *expected* citation rate (*MECR*) as the average of papers of the same datatype and publication year within a reference set (representing, for example, a field of science). The Relative Citation Rate (= *MOCR/MECR*) is thus normalized with unity as the world average.

The two normalizations—in terms of journals and/or fields, and with reference to a world average—can be combined. The Leiden "crown indicator" (CPP/FCSm), for example, was defined by Moed *et al.* (1995) as the average citation rate of the sample under study (citations per publication: CPP) divided by the mean citation score at the field level (FCSm), formalized as follows:[2]

$$CPP / FCSm = \frac{\sum_{i=1}^{m} observed_i / m}{\sum_{i=1}^{n} expected_i / n} \quad (1)$$

Similarly, the current evaluation standard of the ECOOM center in Louvain (Belgium) uses the Normalized Mean Citation Rate (*NMCR*; Glänzel *et al.*, 2009) based on the same principles, but sometimes using their own classification system of journals for the field delineation (Glänzel & Schubert, 2003; cf. Rafols & Leydesdorff, 2009).

[2] A similar normalization can be performed at the level of each journal (CPP/JCSm) as originally proposed by Schubert & Braun (1986). Vinkler (1986) first proposed to extend this expectation to the field level.

The division of two means results in mathematical inconsistency because the order of operations is violated: according to this rule, one should first divide and then add. Instead, one could normalize as follows:

$$MNCS = \left(\sum_{i=1}^{n} \frac{observed_i}{expected_i}\right) / n \qquad (2)$$

In reaction to this critique by Opthof & Leydesdorff (2010; cf. Lundberg, 2007), the Centre for Science and Technology Studies (CWTS) in Leiden has changed the crown indicator (Waltman *et al.*, 2011a), but not all other centers have followed suit. CWTS called this "new crown indicator" the Mean Normalized Citation Score or *MNCS*. One advantage of this indicator is that the mean is a statistics with a standard deviation, and consequently a standard error for the measurement can be defined. Waltman *et al.* (2011b) have shown that this new indicator is mathematically consistent (cf. Egghe, 2012; Vinkler, 2012).

In the sideline of these debates about citation indicators, two further steps were taken. First, Leydesdorff & Opthof (2011) proposed abandoning the idea of journal classification, instead using the *citing* articles as the reference set of the *cited* articles, and then normalizing by counting each citation in proportion to the length of the reference list in the citing paper (1/*NRef*; the field *NRef* is available in the *Science Citation Index*).

Differences in citation behaviour among fields of science can thus be normalized at the source. This fractional counting of citations was previously used by Small & Sweeney (1985) for the mapping of co-citations; and "source-normalization" has been proposed by Zitt & Small (2008) and Moed (2010), albeit in other contexts (Leydesdorff & Opthof, 2010).

Leydesdorff & Bornmann (2011a) could show that the differences in impact factors among fields were no longer statistically significant when using such fractional counting instead of counting each citation as a full point. Another advantage of this methodology is the possibility to evaluate interdisciplinary units fractionally across different disciplines. Zhou & Leydesdorff (2011), for example, have shown that the department of Chinese Language and Literature of the Tsinghua University in Beijing was upgraded in the evaluation from 19$^{th}$ to 2$^{nd}$ position by using fractional counting (given the scarcity of journal citations in the humanities; Leydesdorff *et al.*, 2011a; Nederhof, 2006). In an evaluation of interdisciplinary programs in innovation studies, Rafols *et al.* (in press) found fractional counting of the citations to improve the measurement greatly, particularly if the set was limited to documents with more than ten references (cf. Leydesdorff, 2012; Leydesdorff, Zhou, & Bornmann, in preparation).

In a second contribution to this controversy, Bornmann & Mutz (2011) proposed using percentile ranks instead of averages, in accordance with the standard practice of the *Science & Engineering Indicators* of the US National Science Board (NSB, 2010). Six percentile ranks are distinguished: the top-1%, top-5%, top-10%, top-25%, top-50%, and

bottom-50%. If a one is assigned to a paper in the bottom category, and a six to a paper in the top-1%, it can be shown that the random expectation is a score of 1.91.

The teams of Bornmann & Mutz and Leydesdorff & Opthof thereafter joined forces and evaluated the seven principal investigator (PIs) of the set that originally triggered this debate (Opthof & Leydesdorff, 2010), but using percentile ranks, and then compared the results with those obtained by using the full range of hundred percentiles. In this study, they proposed one further step: one should distinguish evaluation of the seven document sets (of publications) as independent samples—which is the current practice—from their evaluation as subsamples of a reference set.

Among the seven PIs, for example, the one ranked highest had 23 papers of which three were in the top-1%. Each of these three papers contributes (1/23) * 6 = 0.261 points to the total score of this PI (PI1). However, PI6 had 65 papers; a single paper in the top-1% would in this case add (1/65) * 6 = 0.092 points to his/her total score. In other words, one should no longer consider these samples as independent, but instead use the grand sum of the total set ($N$ = 248) so that each PI is credited equally with (1/248) * 6 = 0.024 points for each paper in the top-1% category. By using this measure, PI6 was upgraded to first position, and PI1 downgraded to fifth in the ranking (Leydesdorff *et al.*, 2011b, Table 1, at p. 1375).

The reference set thus provides a standard, and each subset is evaluated as a part of this superset. This reference set can be any encompassing superset, including, for example, a definition of the field in terms of relevant journals or specific keywords (e.g., Bonaccorsi & Vargas, 2011). The crucial point is that each paper adds to the impact score proportionally, that is, according to its percentile rank. Impact is thus no longer dependent on dividing by the number of papers in a specific subset: if two researchers have an equal number of papers in the highest category, adding more papers to one of the sets will add to its percentage impact even if the added papers are in a lower category. In the case of averaging (or using the median) these additional papers would decrease the impact, and thus "punish" productivity.

Leydesdorff & Bornmann (2011b) have elaborated this approach into an indicator: the *Integrated Impact Indicator* (*I3*). In the meantime, a website is available at http://www.leydesdorff.net/software/i3 which provides the relevant routines. The indicator can formally be written as follows: $I3 = \sum_i x_i * n(x_i)$, in which $x_i$ denotes the percentile (rank) value *i*, and *n* the number of papers with this value. *I3* leaves the parametric domain of averages behind and moves to non-parametric statistics using percentiles. The percentiles can be computed as a continuous random variable. Aggregation of the values in the six percentile rank classes distinguished above, leads to an indicator (*PR6*) that weights highly-cited papers more than lower-cited ones. I shall use both these indicators throughout this study.

Using *I3* or *PR6*, one can test whether a contribution is statistically significant above or below expectation using the *z*-test, and one can also compare citation distributions in terms of significant differences using Dunn's test. These are two distinct questions which should not be confused. For example, the citation distribution of the *Proceedings of the National Academy of Science of the USA* and the *Proceedings of the Estonian Academy of Science* are not significantly different, but the impact of the former is significantly above the expected impact using *I3*, while this is not the case for the latter (Leydesdorff & Bornmann, 2011a).

The concept of impact has thus been redefined. Using averages (as in the case of impact factors), the *N* in the denominator has an adverse effect on productivity. When a researcher coauthors with his PhD students, the average citation rate of the group is likely to go down, but the total impact of this research project increases. As Bensman (2008) noted, Garfield (e.g., 1972, 1979b) chose to normalize the ISI-impact factor by dividing by *N* in order to prevent the larger journals from overshadowing the smaller. Bensman then proposed using "Total Citations" instead as an indicator of reputation, because the latter correlated more clearly with the subjective appreciation of faculty than the Impact Factors (IFs). Our percentile-rank approach, however, appreciates that highly-cited papers should be weighted more than less frequently cited ones.

The division by *N* was just too coarse: using an average assumes a normal distribution, whereas the skewness in different citation distributions can be made comparable when normalized on the 1–to–100 scale of percentiles. How the percentiles are aggregated, for example, into the six percentile ranks of the NSF used by Bornmann & Mutz (2011), or in terms of quartiles, is a normative decision which can be made independently from the technical analysis in terms of hundred percentiles. In what follows, I use both the scheme of six percentile ranks and the baseline of 100 percentiles to evaluate the case of journals subsumed under the ISI Subject Category of "Nanoscience & nanotechnology." In order to make my results comparable with the IFs, I use whole counts and two years of cited data (2007 and 2008) and compare with the IF-2009 (the most recent one available at the time of this research; Spring 2011). As noted, *I3* is defined more abstractly; it can be used with any reference set and also with fractional counts.

## The "nano-set" in the *Science Citation Index*

The ISI Subject Category "Nanoscience & nanotechnology" (NS) entered the SCI in 2005 with 27 journals and more than doubled to include 59 journals by 2009. One of these journals (*ACM Applied Materials & Interfaces*) was added to the database only in 2009, and consequently no records for the two previous years were included (as of the time of this study). I limit the set to the 31,644 citable items in these 58 journals during the years 2007 and 2008 (that is: 25,271 articles, 5,488 proceedings papers, 709 reviews, and 176 letters). The percentile of each paper is determined with reference to the set with the same publication year and document type, respectively.

A simple counting rule for percentiles is the number of papers with lower citation rates divided by the total number of similar records in the set. The resulting value can be rounded into percentiles or used as a continuous random variable (so-called "quantiles"). The advantage of this counting rule over other possible ones is that tied ranks are accorded their highest possible values. Other rules are also possible: Pudovkin & Garfield (2009), for example, first averaged tied ranks.

Before turning to the methodological details, let me first specify the problem by taking the journal with the highest impact factor in this category—that is, *Nature Nanotechnology* with IF-2009 = 26,309—and comparing it with the third in rank: *Nano Letters* with IF-2009 = 9.991.[3]

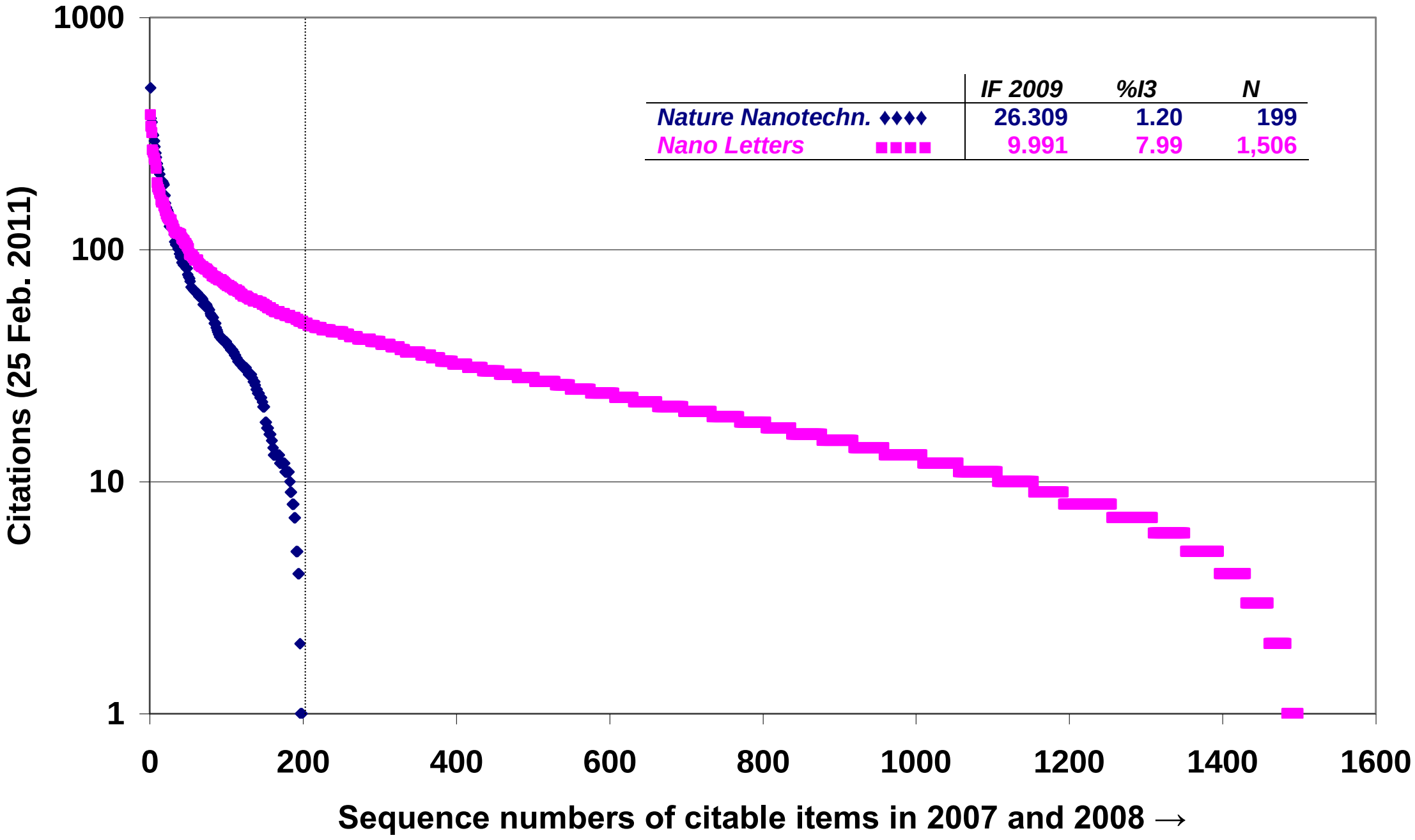

**Figure 1**: Citation curves for *Nature Nanotechnology* (n = 199 publications) and *Nano Letters* (n = 1,506).

Figure 1 shows that the lower impact factor of *Nano Letters* is entirely due to the large tail of the distribution. In the left part of this figure (the 199 most-highly cited papers), the average citation score of *Nano Letters* (89.78 ± 0.27) is higher than that of *Nature Nanotechnology* (66.45 ± 0.38).

[3] Second in rank in 2009 was the journal *Nano Today*, with IF = 13.237.

Analogously, using an average value would underestimate productivity of a university group because of the *N* in the denominator. If one adds two groups together or merges these two journals (in a thought experiment), the impact of the merged group should, in my opinion, be not the average of the two previous groups, but their sumtotal. However, this summation needs to be qualified: papers in the top-1% range are to be added to other papers in the top-1%, etc. *mutatis mutandis*. If we compare, thus qualified, the six percentile ranks for these two journals, the results can be seen in Figure 2:

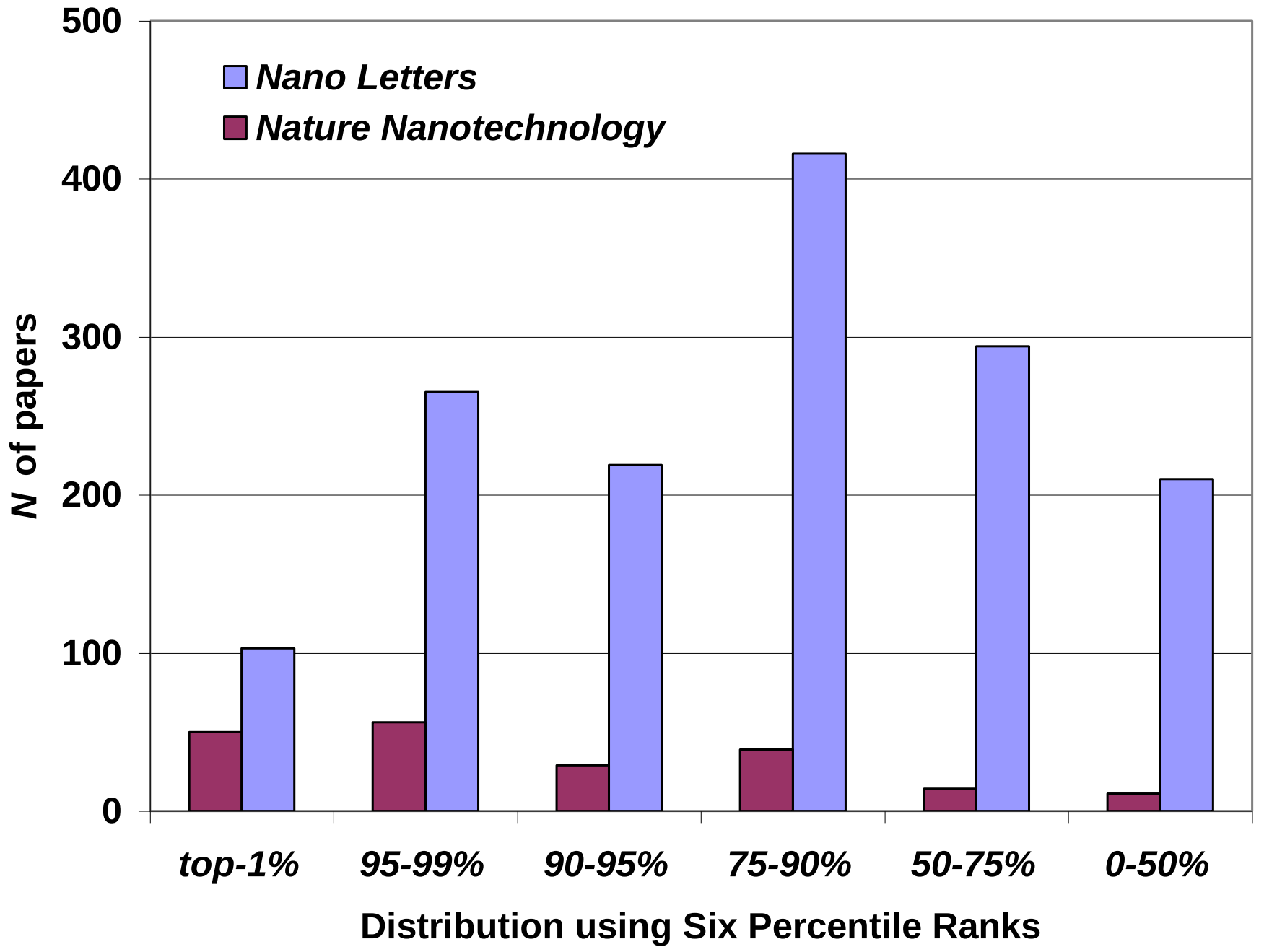

**Figure 2**: Distribution of six percentile ranks of publications in *Nano Letters* and *Nature Nanotechnology* with reference to the 58 journals of the ISI Subject Category "nanoscience & nanotechnology."

Figure 2 shows that *Nano Letters* outperforms *Nature Nanotechnology* in all six classes. However, if one divides the total scores by the number of publications in each category, *Nano Letters* becomes the less important journal (in all categories). Thus one should not divide, but integrate: impacts adds up. The impact of two collisions is the (vector) sum of the two momenta. The use of the word "impact factor" to denote an average value has confused the semantics for decades (Sher & Garfield, 1965; Garfield, 1972).

The integrals in this stepwise function are equal to $\sum_i x_i * f(x_i)$ in which *x* represents the percentile rank and $f$(x) the frequency of that rank. For example, $i = 6$ in the case above, or $i = 100$ when using all percentiles as classes. The hundred percentiles can be considered as a continuous random variable, and these "quantiles" form the baseline. Note that the percentile is thereafter a characteristic of each individual paper. Thus, different aggregations of papers are possible, for example in terms of journals, nations, institutes, cities, or authors. The function integrates both the number of papers (the "mass") and their respective quality in terms of being-cited, but after normalization represent them as percentiles with reference to a set.

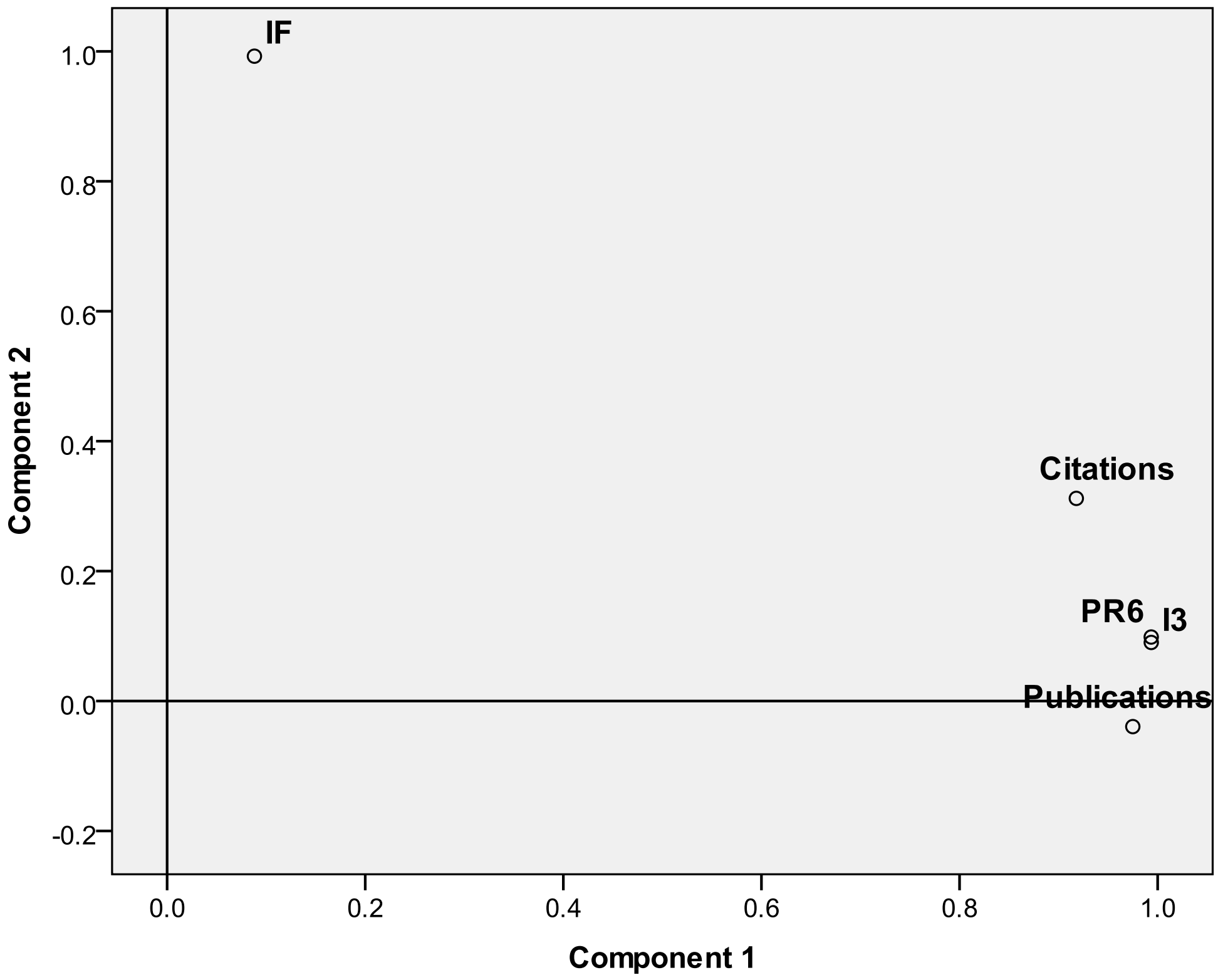

**Figure 3**: Factor analysis (varimax rotated) of the number of publications, citations, *IF*s, *I3*-values (using 100 percentiles), and values using 6 percentile ranks (*PR6*); the 58 journals attributed to the ISI Subject Category "Nanoscience & nanotechnology" are the cases.

Figure 3 shows the results of a factor analysis using the 58 journals attributed to the ISI Subject Category "Nanoscience & nanotechnology" as cases, and the numbers of publications, citations, *IF*s, *I3* values for both 100 and for 6 classes (*PR6*). The citations and publications correlate highly ($r = 0.835$, $p \leq .01$) because citations correlate with size (Bensman, 2007). In my opinion, an impact indicator can only be meaningful if it also correlates with these indicators of productivity and quality. But this is not the case for the impact factor, because it is based on dividing these two indicators.

*I3* correlates with both citations ($r = .935$, $p \leq .01$) and publications ($r = .963$, $p \leq .01$) more than these two values correlate mutually: this then takes both productivity and quality into account. *PR6* is slightly more sensitive to high citation rates than *I3* (based on hundred percentiles), but the difference between the two indicators is not significant.

## Methods and materials

The 58 journals in the ISI Subject Category "Nanoscience & nanotechnology" conveniently contain a large sample: 31,644 citable items, with 65,173 addresses, and 146,432 authors. All records were downloaded on February 25, 2011. Dedicated software was developed and brought online at http://www.leydesdorff.net/software/i3. The routines add the values for the two relevant indicators (*I3* and *PR6*) at the article level to a set downloaded from the Web-of-Science (WoS) and organized into relational databases. The percentiles can be summed into categories using the various possible schemes for an evaluation. As noted, I use the six ranks of the NSF as an example in addition to the quantiles (between 0 and 100) as a continuous random variable.

The simple counting rule for quantiles—specified above—may generate problems when a reference set is smaller than 100. For example, if a journal includes among its publications only 10 reviews each year, the highest possible percentile would be the 90%—nine out of ten—whereas this could be the $99^{th}$ percentile. Rüdiger Mutz (*personal communication*, February 18, 2011) suggested adding 0.9 to each percentile value, which solved part of the problem (Leydesdorff & Bornmann, 2011b). Rousseau (2012) proposed adopting as a counting rule not "less than" (<), but "less than or equal to" (≤) the item under study. All papers would then have a chance to be in the top rank of the 100th percentile, but the resulting values are slightly higher. I did not apply this solution in the current study (Leydesdorff & Bornmann, in press). Most recently, Schreiber (in press) proposed another solution to this problem.

The file of 31,644 citable items each with a value for "times cited" ("TC" provided by the database), *I3* (for the hundred percentiles), and *PR6* (for the six classes of the NSF) can be imported into SPSS (v.18); and then the routine "Compare Means," using the journals—countries, cities, etc., respectively—in each set as the independent (grouping) variable, can be used for determining all relevant statistics. (This routine also allows for bootstrapping and the determination of confidence levels.) The *I3*-values and *PR6*-values are based on summation for the units of analysis under evaluation. The SEM (standard error of measurement) values are based on averaging and will not be used below, but are also available from this routine.

The *I3* and *PR6* values can be recomposed by aggregating for each subset—for example, each journal or a subset with a specific institutional address—and expressed as a percentage of the total *I3* (or *PR6*) value for the set. The routine isi2i3.exe also provides tables at the level of authors, institutes,[4] countries, and journals, with aggregates of these variables. (Alternatively, one can use the procedure "Aggregate cases" in SPSS or generate pivot tables in Excel.) The absolute numbers can be expressed as percentages of the set which can then be compared and used for ranking. Examples will be provided in the results section below.

---

[4] The aggregation at the city level presumes processing the address field using i3cit1.exe, available at http://www.leydesdorff.net/software/i3.

Both *I3* and *PR6* are size-dependent because a unit with 1000 publications has ten times more chance to have one in the top-1% range than a unit with only 100 publications. These indicator values can be tested against the publication rates: *ex ante*, all publications have an equal chance to be cited. One can use the *z*-test for two independent populations to test observed versus expected rates for their statistical significance (Sheskin, 2011, p. 656).[5] Note that the impact *I3* is not independent of the number of publications. I shall return to this issue below.

Citation curves (such as in Figure 1 above) can be tested against each other using "Multiple Comparisons" with Bonferroni correction in SPSS. When the confidence levels do not intersect, the distributions are flagged as significantly different. In the case of non-parametric statistics, Dunn's test can be simulated by using LSD ("least significant differences") with family-wise correction for Type-I error probability (Levine, 1991, at pp. 68 ff.).[6] The routine in SPSS is limited to a maximum of 50 comparisons. Alternatively, one can use the Mann-Wihitney U test with the same Bonferroni correction between each individual pair.

Journals (etc.) with significantly different or similar citation patterns, can be visualized as a graph in which homogenous sets are connected while significantly different nodes are not. I shall use the algorithm of Kamada & Kawai (1989) in Pajek for this visualization. The *k*-core set which are most homogenous in terms of citation distributions can thus be visualized. Analogously, other variables which are attributed at the article level, such as the percentile values *I3* and *PR6*, can be analyzed. Differences and similarities in citation distributions can be expected to change after normalization in terms of percentile values. The latter exhibit differences in impact (*I3* and/or *RP6*).

[5] Leydesdorff & Bornmann (2011) mention this option, but use the standardized residuals which can also be expected to be chi-square distributed. Unlike the *z*-test, however, this assumption is not yet mathematically proven in the literature.

[6] This error probability increases with the number of pairwise comparisons *c*: $c = n * (n - 1) / 2$. For 50 cases (e.g., journals), the number of pairwise comparsions is 50 * 49 / 2 = 1,225, and one should therefore test at a significance level of 0.05/ 1,225 = 0.000041 instead of $p \leq 0.05$.

In summary, one can assess each percentage impact (with reference to a set) in terms of whether the contribution (expressed as a percentage impact of the set) is significantly above or below the expectation. In accordance with the convention in SPSS but using plus and minus signs, I use double $^{++}$ for above and $^{--}$ for below expectation at the 1% level of significance testing, and single $^{+}$ and $^{-}$ at the 5% level. The significance of the *contribution* is a different question from whether its ranking is based on significant *differences* in the underlying distributions of the citations or *I3*-values. The latter analysis enables us to indicate groups which may differ in size but otherwise be homogenous.

## Results

### *Journal Evaluation*

Table 1 provides the rankings of the 20 nano-journals most highly ranked in terms of their impact measured as *I3* (column b). As expected, the relation with the IFs is virtually absent ($r$ = .178; *n.s.*). *Nano Letters* ranks second on this list in terms of its IF, with *Nature Nanotechnology* in first place. But while *Nano Letters* ranks third from the top in terms of *I3, Nature Nanotechnology* in this case occupies only the 19$^{th}$ position.

**Table 1**: Rankings between 20 journals of "Nanoscience & nanotechnology" with highest values on *I3* (expressed as percentages of the sum) compared with *IF*s, total citations, and % *I3* (*6PR*).

| | Journal | N of papers (a) | % I3 (b) | IF 2009 (c) | Total citations (d) | % I3 (6PR) (e) |
|---|---|---|---|---|---|---|
| 1 | *J Phys Chem C* | 5,585 [1] | 20.71 [1] $^{++}$ | 4.224 [9] | 62,072 | 18.79 [1] $^{++}$ |
| 2 | *Mater Sci Eng A-Struct Mater* | 3,222 [2] | 8.82 [2] $^{--}$ | 1.901 [12] | 15,920 | 8.64 [2] $^{--}$ |
| 3 | ***Nano Lett*** | 1,507 [6] | 7.99 [3] $^{++}$ | 9.991 [2] | 42,159 | 8.38 [3] $^{++}$ |
| 4 | *Nanotechnol* | 2,656 [3] | 7.96 [4] $^{--}$ | 3.137 [10] | 18,748 | 7.22 [5] $^{--}$ |
| 5 | *Advan Mater* | 1,512 [5] | 7.61 [5] $^{++}$ | 8.379 [3] | 34,581 | 7.59 [4] $^{++}$ |
| 6 | *Adv Funct Mater* | 883 | 4.24 [6] $^{++}$ | 6.990 [5] | 16,629 | 4.04 [7] $^{++}$ |
| 7 | *J Nanosci Nanotechnol* | 1,699 [4] | 3.74 [7] $^{--}$ | 1.435 [15] | 5,946 | 4.07 [6] $^{--}$ |
| 8 | *Biosens Bioelectron* | 821 | 3.57 [8] $^{++}$ | 5.429 [8] | 11,583 | 3.34 [8] $^{++}$ |
| 9 | *Scripta Mater* | 1,153 | 3.43 [9] $^{--}$ | 2.949 [11] | 8,656 | 3.13 [11] $^{--}$ |
| 10 | *Microporous Mesoporous Mat* | 1,041 | 3.40 [10] $^{++}$ | 2.652 [12] | 7,140 | 3.14 [10] |
| 11 | *Microelectron Eng* | 1,064 | 3.14 [11] $^{--}$ | 1.488 [13] | 4,090 | 3.17 [9] $^{-}$ |
| 12 | *Small* | 595 | 2.58 [12] $^{++}$ | 6.171 [7] | 9,811 | 2.40 [14] $^{++}$ |
| 13 | *Lab Chip* | 520 | 2.34 [13] $^{++}$ | 6.342 [6] | 8,930 | 2.21 [15] $^{++}$ |
| 14 | *J Vac Sci Technol B* | 962 | 2.29 [14] $^{--}$ | 1.460 [14] | 3,471 | 2.43 [13] $^{--}$ |
| 15 | *Physica E* | 1,004 | 2.26 [15] $^{--}$ | 1.177 [16] | 3,126 | 2.64 [12] $^{--}$ |
| 16 | *J Micromechanic Microengineer* | 778 | 1.86 [16] $^{--}$ | 0.541 [19] | 186 | 1.86 [16] $^{--}$ |
| 17 | *Acs Nano* | 349 | 1.73 [17] $^{++}$ | 7.493 [4] | 6,845 | 1.72 [17] $^{++}$ |
| 18 | *Microelectron Rel* | 627 | 1.20 [18] $^{--}$ | 1.117 [17] | 1,767 | 1.44 [19] $^{--}$ |
| 19 | ***Nat Nanotechnol*** | 199 | 1.20 [19] $^{++}$ | 26.309 [1] | 13,224 | 1.47 [18] $^{++}$ |
| 20 | *Microsyst Technol* | 475 | 1.00 [20] $^{--}$ | 1.025 [18] | 1,265 | 1.17 [20] $^{--}$ |

Note. $^{++}p < 0.01$ above the expectation; $^{+}p < 0.05$ above the expectation;
$^{--}p < 0.01$ below the expectation; $^{-}p < 0.05$ below the expectation.
Ranks are added between brackets.

Note that *Nano Letters* ranks only sixth in terms of the number of publications (column a), but third in terms of the impact *I3* and *6PR* (columns b and e, respectively). Its impact is thus (significantly!) above expectation. *Mater Sci Eng A-Struct Mater*,[7] in second position, is larger in terms of numbers of publications, but relatively low in terms of citations. Therefore, it scores significantly below the expected citation rates in columns (b) and (e). Analogously, the c/p ratio for *Nature Nanotechnology* (13,244/199 =) 66.45 corresponds to its high IF, but, as noted, this is entirely due to the low number in the denominator ($N$ = 199). The *I3* and *PR6* values are determined by both productivity and citation rates. In this case, values for *I3* and *PR6* were virtually identical ($r$ = .998; $\rho$ = .966; $p \leq 0.01$; $N$ = 58).

[7] The full title of this journal is: *Materials Science and Engineering A-Structural Materials Properties Microstructure and Processing.*

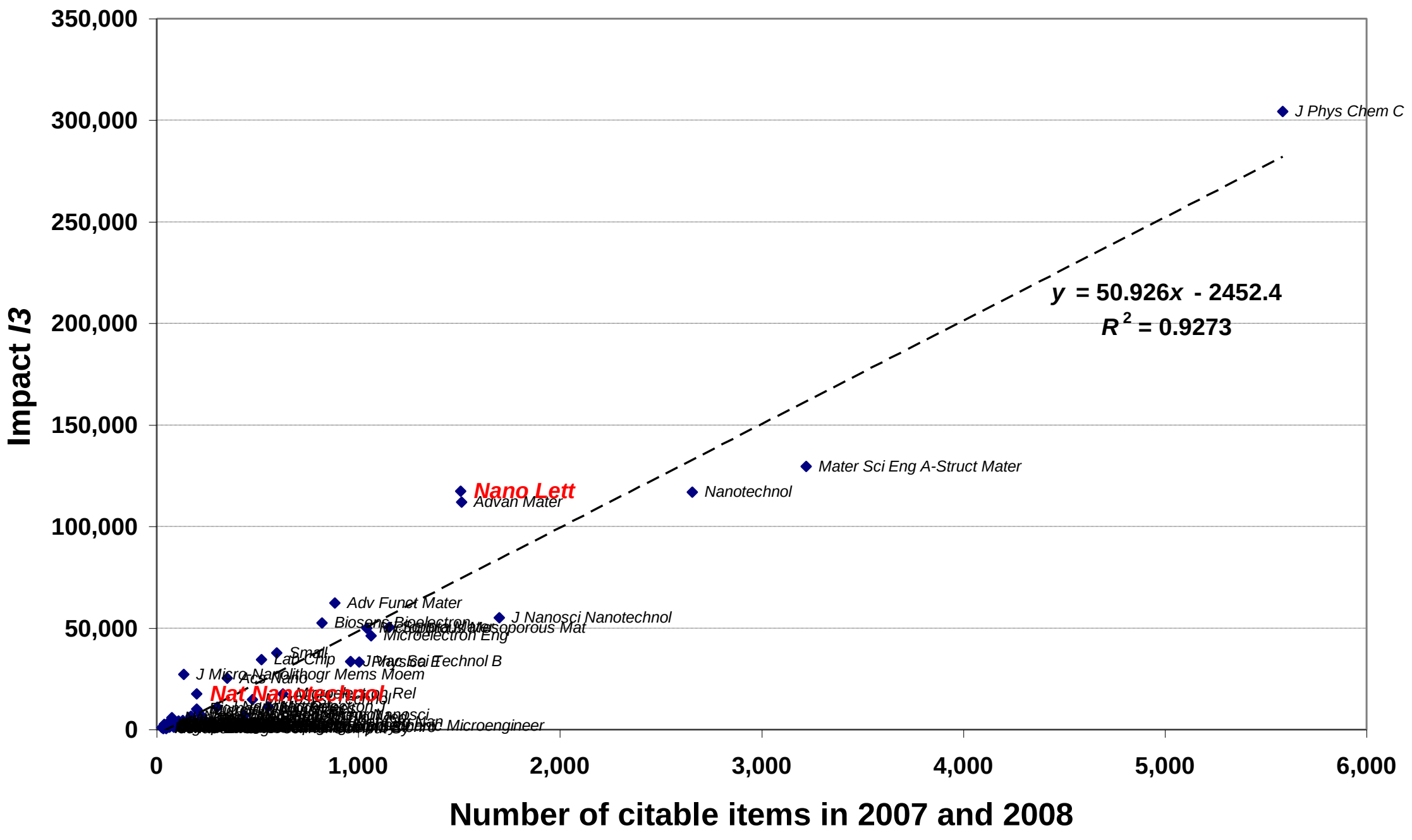

**Figure 4**: Regression of impact (*I3*) against the number of citable publications in 2007 and 2008.

Figure 4 shows the relation between impact (*I3*) and size. Both *Nano Letters* and *Nature Nanotechnology* (indicated in red) lie above the regression line, whereas the journal *Mater Sci Eng A-Struct Mater* does not. As shown in Table 1**,** these deviations from the respective expectations are highly significant ($p < 0.01$). The curve also explicates that *Nano Letters* is a large journal, whereas *Nature Nanotechnology* is not very different in size and impact from a large number of specialist journals that are much smaller.

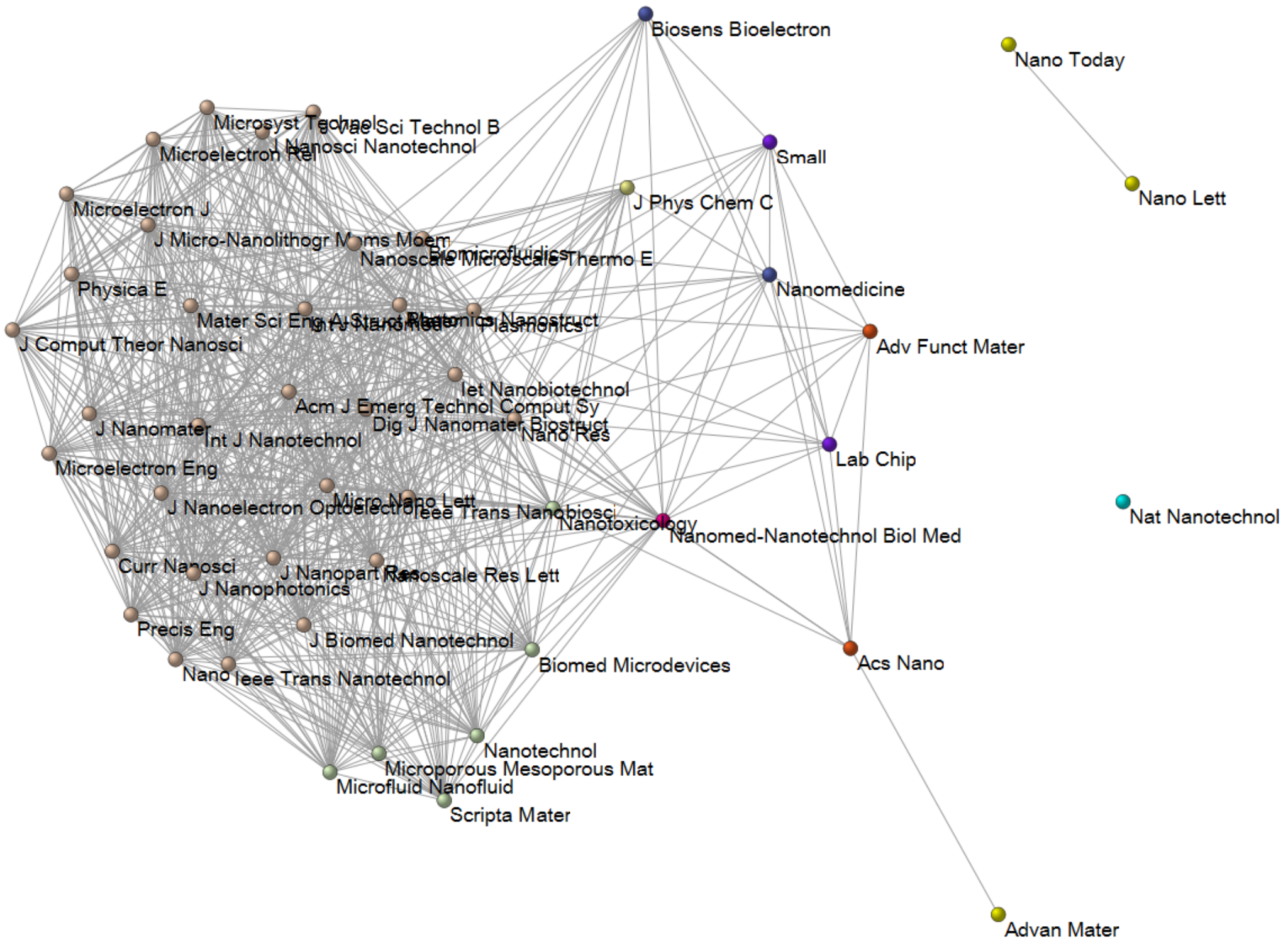

**Figure 5**: Homogenous groups in terms of distribution of citation distributions among 50 journals in "nanoscience & nanotechnology." An edge indicates similarity in the distribution at the 5% level ($p < 0.05/1225$; using Bonferroni correction as explained in footnote 6).

Figure 5 provides the results of the other statistical test, namely Dunn's test for multiple comparisons. As noted, this test is limited in SPSS to 50 cases. I used the 50 journals with the highest IFs. Thirty-two of these journals belong to a core group ($k$ = 29). This means that their citation patterns are not statistically different from at least ($k$ – 1 =) 28 other journals in this group. Seven more journals are statistically homogenous with at least 25 other journals. However, the citation pattern of *Nature Nanotechnology* is completely different from all other journals in the group, while *Nature Letters* shares its citation pattern only with *Nano Today*, the journal with the second largest IF (= 13.237) within this set, but again with a much lower *I3* (of 0.18%).

If we repeat this analysis with the distributions of *I3*—instead of citation distributions across these fifty journals (not shown), all journals are connected in a single graph (24 in a $k = 17$ core with ten more journals connected to this component at the level of $k = 11$). However, the *I3*-distribution of *Nature Nanotechnology* is statistically similar only to the one of *Nano Today*, while this latter journal is connected to five other, among them *Nano Letters*. Not surprisingly, the impact pattern of *Nano Letters* is not otherwise different from *ACS Nano*, another journal of the American Chemical Society. *ACS Nano* was ranked 17$^{th}$ in Table 1, and is another highly specialized journal with a relatively high *IF* (7.493), but a lower value of *I3*.

In summary, *I3* and the *IF* measure very different aspects of impact. Whereas *I3* can be considered as a measure of impact comparable to the *h*-index, but sophisticated in its statistics and elaboration, the *IF* is based on a division by the number of citable items in the denominator. Bensman (2008) noted that Garfield thus wished to distinguish this indicator from “total cites” in order to control for the size of journals. However, the choice of the mean as a measure for impact has unfortunately led to the mistaken assumption that impact can be measured by averaging.

*Evaluation of nations*

Because *I3* is not an average over a distribution, but a summation of values of times cited normalized at the level of individual papers, aggregations other than those in terms of journals are equally possible. Since Narin (1976) and Small & Garfield (1985) one has been inclined to consider the two major aggregations in terms of journals and nations as the orthogonal axes of a matrix for citation analysis; but set can also be aggregated using criteria such as keywords or author groups across journals and/or institutional addresses. Each paper has a percentile rank with reference to the set which represents the relevant field. Each subset can be tested on the significance of the contribution, that is, above or below statistical expectation.

**Table 2**: Rankings among top-20 countries in terms of *I3* (expressed as percentages of the sum) compared with number of publications, total citations, and % *I3*(*6PR*) for the 58 journals in the ISI Subject Category "Nanoscience & nanotechnology."

| | | *Publications* | *Citations* | *Cit/Pub* | *% I3* | *% PR6* |
|---|---|---|---|---|---|---|
| 1 | USA | 16,039 | 234,370 | 14.61 | $27.73^{++}$ | $27.93^{++}$ |
| 2 | Peoples R China | 8,653 | 83,079 | 9.60 | $12.88^{--}$ | $12.74^{--}$ |
| 3 | Japan | 4,798 | 39,469 | 8.23 | $7.08^{--}$ | $7.00^{--}$ |
| 4 | Germany | 3,802 | 43,231 | 11.37 | $6.54^{++}$ | $6.46^{++}$ |
| 5 | South Korea | 4,250 | 32,772 | 7.71 | $6.10^{--}$ | $6.14^{--}$ |
| 6 | France | 2,803 | 21,099 | 7.53 | $4.04^{--}$ | $3.92^{--}$ |
| 7 | England | 1,963 | 24,875 | 12.67 | $3.24^{++}$ | $3.24^{++}$ |
| 8 | Taiwan | 2,529 | 19,068 | 7.54 | $3.24^{--}$ | $3.30^{--}$ |
| 9 | Italy | 2,021 | 15,738 | 7.79 | $3.00^{--}$ | $2.92^{--}$ |
| 10 | India | 2,215 | 13,593 | 6.14 | $2.67^{--}$ | $2.72^{--}$ |
| 11 | Spain | 1,527 | 14,692 | 9.62 | $2.45^{++}$ | $2.33^{-}$ |
| 12 | Canada | 1,360 | 13,418 | 9.87 | 2.09 | 2.04 |
| 13 | Singapore | 1,219 | 13,785 | 11.31 | $1.94^{++}$ | 1.91 |
| 14 | Netherlands | 872 | 10,951 | 12.56 | $1.53^{++}$ | $1.51^{++}$ |
| 15 | Australia | 919 | 9,560 | 10.40 | $1.51^{++}$ | 1.45 |
| 16 | Switzerland | 682 | 9,565 | 14.02 | $1.29^{++}$ | $1.27^{++}$ |
| 17 | Belgium | 668 | 5,683 | 8.51 | $1.05^{++}$ | 1.04 |
| 18 | Sweden | 566 | 6,661 | 11.77 | $0.96^{++}$ | $0.96^{++}$ |
| 19 | Israel | 525 | 5,662 | 10.78 | $0.86^{++}$ | 0.78 |
| 20 | Russia | 1,023 | 3,410 | 3.33 | $0.81^{--}$ | $1.11^{--}$ |

Note. $^{++}p < 0.01$ above the expectation; $^{+}p < 0.05$ above the expectation;
$^{--}p < 0.01$ below the expectation; $^{-}p < 0.05$ below the expectation.

Table 2 ranks the 20 leading nations (based on "integer counting" attributing a full point to each address).[8] Table 3 disaggregates further to the level of institutions. Although the contributions of the Asian countries are evaluated in Table 2 as significantly below expectation (with the exception of Singapore), Table 3 informs us that individual institutions in these countries may perform significantly above expectation in terms of their impact. The Chinese Academy of Sciences in Beijing leads the list with over 60% more impact than the second and third (American) universities. Except for the National University of Singapore, the contributions of non-Chinese centers in Asia rank as less significant or even (sometimes significantly) below expectation.

[8] Alternatively, "fractional counting" attributes addresses proportionally to the number of addresses/paper.

**Table 3**: Rankings among the 20 top-institutes in terms of *I3* (expressed as percentages of the sum) compared with number of publications and citations, and % *I3*(*6PR*) for the 58 journals in the ISI Subject Category "Nanoscience & nanotechnology."

| | | *Publications* | *Citations* | *c/p* | *% I3* | *% PR6* |
|---|---|---|---|---|---|---|
| 1 | Chinese Acad Sci, Beijing, Peoples R China | 824 | 11,258 | 13.66 | 1.43$^{++}$ | 1.42$^{++}$ |
| 2 | MIT, Cambridge, USA | 420 | 8,372 | 19.93 | 0.89$^{++}$ | 0.90$^{++}$ |
| 3 | Univ Calif Berkeley, Berkeley, USA | 425 | 8,639 | 20.33 | 0.88$^{++}$ | 0.92$^{++}$ |
| 4 | Natl Univ Singapore, Singapore, Singapore | 470 | 6,483 | 13.79 | 0.83$^{++}$ | 0.82$^{++}$ |
| 5 | Northwestern Univ, Evanston, USA | 304 | 7,594 | 24.98 | 0.63$^{++}$ | 0.68$^{++}$ |
| 6 | Univ Michigan, Ann Arbor, USA | 301 | 5,633 | 18.71 | 0.63$^{++}$ | 0.63$^{++}$ |
| 7 | Univ Illinois, Urbana, USA | 307 | 6,905 | 22.49 | 0.61$^{++}$ | 0.62$^{++}$ |
| 8 | Georgia Inst Technol, Atlanta, USA | 307 | 5,658 | 18.43 | 0.61$^{++}$ | 0.62$^{++}$ |
| 9 | Nanyang Technol Univ, Singapore, Singapore | 383 | 4,110 | 10.73 | 0.60$^{+}$ | 0.59 |
| 10 | Penn State Univ, University Park, USA | 293 | 4,519 | 15.42 | 0.52$^{++}$ | 0.53$^{++}$ |
| 11 | Univ Cambridge, Cambridge, UK | 276 | 4,677 | 16.95 | 0.52$^{++}$ | 0.52$^{++}$ |
| 12 | Seoul Natl Univ, Seoul, South Korea | 314 | 4,135 | 13.17 | 0.51$^{++}$ | 0.51$^{+}$ |
| 13 | Peking Univ, Beijing, Peoples R China | 285 | 3,698 | 12.98 | 0.50$^{++}$ | 0.50$^{++}$ |
| 14 | Rice Univ, Houston, USA | 239 | 6,170 | 25.82 | 0.49$^{++}$ | 0.55$^{++}$ |
| 15 | Purdue Univ, Lafayette, USA | 282 | 3,931 | 13.94 | 0.48$^{++}$ | 0.48$^{++}$ |
| 16 | Tohoku Univ, Miyagi, Japan | 349 | 2,274 | 6.52 | 0.47$^{--}$ | 0.47$^{--}$ |
| 17 | Univ Washington, Seattle, USA | 210 | 5,126 | 24.41 | 0.47$^{++}$ | 0.50$^{++}$ |
| 18 | Natl Inst Mat Sci, Ibaraki, Japan | 271 | 2,831 | 10.45 | 0.46$^{++}$ | 0.44 |
| 19 | Natl Cheng Kung Univ, Tainan, Taiwan | 340 | 2,256 | 6.64 | 0.43$^{--}$ | 0.43$^{--}$ |
| 20 | Stanford Univ, Stanford, USA | 200 | 5,894 | 29.47 | 0.42$^{++}$ | 0.47$^{++}$ |

Note. $^{++}p < 0.01$ above the expectation; $^{+}p < 0.05$ above the expectation;
$^{--}p < 0.01$ below the expectation; $^{-}p < 0.05$ below the expectation.

*More specific delineation of the nano-set*

Because of its interdisciplinarity, the ISI Subject Category "Nanoscience & nanotechnology" can be considered as a mixed bag (Leydesdorff & Rafols, 2012; Rafols *et al.*, 2010). Factor analysis of the citation matrix of these 58 journals in terms of the cited patterns provided us with a first factor showing robustly the same 15 journals with factor loadings > .4 using both four or five factors for the extraction. This Factor One can be considered as the representation of a more homogenous set of journals which publish about nanoscience and nanotechnology from the perspective of condensed matter physics and chemistry-based nanotechnology. Other factors are focused on micro- and nanofluidics (factor 2), nano-medicine (factor 3), and nano-electronics (factor 4; Ismael Rafols, *personal communication*, 20 March 2011). Factor One explains appr. 36% of the variance in the matrix of aggregated citation relations among the 58 journals under study.

Let us take this restricted domain of 15 journals (*N* = 14,794 citable items)[9] as a specific and typical research specialty (cf. Huang *et al.*, 2011). Table 4 provides a summary of the contributions of journals, and the rankings of (top) countries, cities, and institutes in this domain. The percentiles were recalculated using these 15 journals as the reference set. In the first column of Table 4, for example, the relative impact of *Advances in Materials* is greater than that of *Nanotechnology*, reversing their ranking (based on 58 journals) as shown in Table 1. Thus, *Nanotechnology* is more highly cited outside this subset than *Advances in Materials*. The indicators of significance are, however, largely the same in both Tables.

[9] The number of addresses is 32,134, and the number of authors: 73,342.

**Table 4**: Integrated impact (%*I3*) of journals and rankings of (15 top) countries, cities, and institutes in the domain of the 15 journals (in column a) more specifically selected as condensed-matter and chemistry-based nanotechnology.

| | *Journals* (*N* = 15)<br>(a) | *Countries* (*N* = 95)<br>(b) | *Cities* (*N* = 1,694)<br>(c) | *Institutes* (*N* = 5,726)<br>(d) |
|---|---|---|---|---|
| 1 | *J Phys Chem C* $^{--}$ | USA $^{++}$ | Beijing, Peoples R China $^{++}$ | Chinese Acad Sci, Beijing, Peoples R China $^{++}$ |
| 2 | *Nano Lett* $^{++}$ | Peoples R China $^{--}$ | Singapore, Singapore | Univ Calif Berkeley, Berkeley, USA $^{++}$ |
| 3 | *Advan Mater* $^{++}$ | Germany $^{++}$ | Seoul, South Korea $^{--}$ | MIT, Cambridge, USA $^{++}$ |
| 4 | *Nanotechnology* $^{--}$ | Japan $^{--}$ | Cambridge, USA $^{++}$ | Northwestern Univ, Evanston, USA $^{++}$ |
| 5 | *Adv Funct Mater* $^{++}$ | South Korea | Berkeley, USA $^{++}$ | Natl Univ Singapore, Singapore, Singapore $^{++}$ |
| 6 | *Physica E* $^{--}$ | UK | Shanghai, Peoples R China $^{--}$ | Univ Michigan, Ann Arbor, USA $^{++}$ |
| 7 | *Small* $^{++}$ | France $^{--}$ | Ibaraki, Japan | Georgia Inst Technol, Atlanta, USA $^{++}$ |
| 8 | *ACS Nano* $^{++}$ | Italy $^{--}$ | Tokyo, Japan $^{--}$ | Univ Illinois, Urbana, USA $^{++}$ |
| 9 | *Nat Nanotechnol* $^{++}$ | Taiwan $^{--}$ | Evanston, USA $^{++}$ | Rice Univ, Houston, USA $^{++}$ |
| 10 | *Nano* $^{--}$ | Spain $^{--}$ | Atlanta, USA $^{++}$ | Peking Univ, Beijing, Peoples R China $^{++}$ |
| 11 | *Nano Res* $^{-}$ | Canada $^{--}$ | Houston, USA $^{++}$ | Univ Cambridge, Cambridge, UK $^{++}$ |
| 12 | *Int J Nanotechnol* $^{--}$ | Singapore | Changchun, Peoples R China $^{++}$ | Univ Washington, Seattle, USA $^{++}$ |
| 13 | *Nano Today* $^{++}$ | India $^{--}$ | Taipei, Taiwan | Penn State Univ, University, USA $^{++}$ |
| 14 | *J Nanomater* $^{--}$ | Australia $^{--}$ | Hong Kong, Peoples R China | Nanyang Technol Univ, Singapore, Singapore $^{-}$ |
| 15 | *Plasmonics* $^{--}$ | Netherlands $^{++}$ | Ann Arbor, USA $^{++}$ | Univ Calif Los Angeles, Los Angeles, USA $^{++}$ |

Note. $^{++}p < 0.01$ above the expectation; $^{+}p < 0.05$ above the expectation;
$^{--}p < 0.01$ below the expectation; $^{-}p < 0.05$ below the expectation.

The table confirms the impression obtained above: Asian units are more concentrated in large metropoles; at the higher level of aggregation of countries the USA is still dominant; and China performs below expectation because of lower citation rates. Except for the UK (University of Cambridge), European institutes and cities do not appear to play a leading role. Japanese cities and centers tend to score below expectation, but several Chinese centers are leading along with American ones.

One can use Dunn's test (as above in Figure 4) for the evaluation, for example, of the extent to which the impact-profiles of the 15 institutions are significantly different and/or homogenous. The I3-distributions of ten American universities and the University of Cambridge in the UK are statistically homogenous. Three of these universities (Rice University, Penn State, the University of Cambridge, UK) and the National University of Singapore have stronger roles at the interface with the otherwise differently profiled Chinese universities. In other words, the rank-ordering among the American-British universities in this core set is also an effect of size, as it is among the Asian ones. If we use citations instead of *I3* values for Dunn's test, the results are not different except that the University of California in Los Angeles (number 15 in rank) is no longer included among the core set of American/UK universities. Thus, the analysis in terms of *I3* refines our ability to compare citation distributions involving multiple comparisons.

What happened to Western Europe? With the exception of Cambridge (UK), it is absent from the listing of institutions in Table 4, but Germany, for example, takes the second position in the ranking of nations. Figure 6 cuts the northwestern section of Europe out of a global map (available at http://www.leydesdorff.net/nano2011/nano2011.htm) using this same data and using methods developed previously (Bornmann & Leydesdorff, 2011; Leydesdorff & Persson, 2010). Bornmann & Leydesdorff (2011) used the top-10% as an indicator of excellence, but one can use *I3* values without setting a threshold for indicating whether impact is above or below expecation.

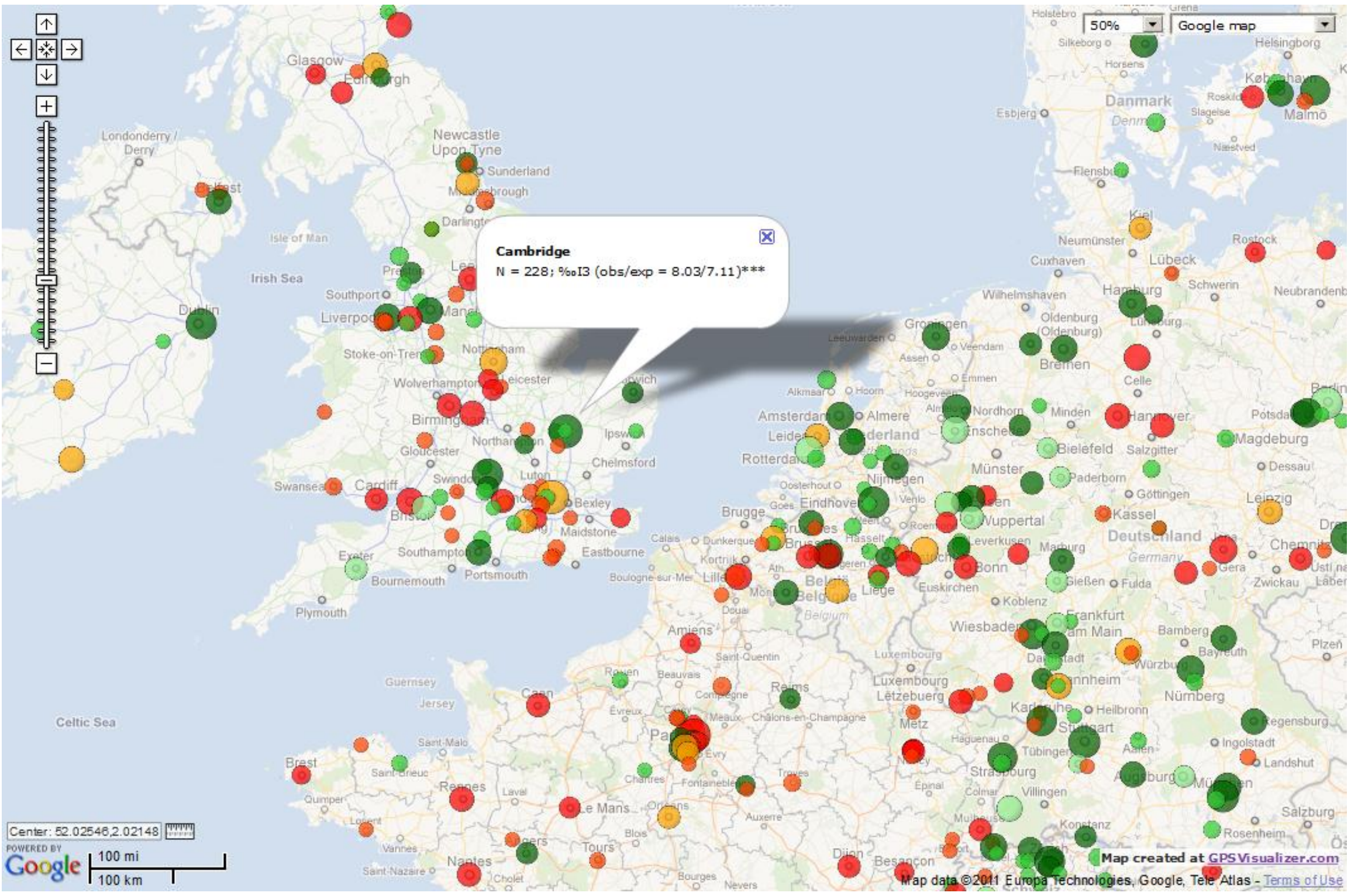

**Figure 6**: Distribution of cities in North-Western Europe above or below expectations in terms of integrated impact on the basis of 15 core journals. The sizes of the nodes are proportional to the logarithm of the number of papers; colors: red indicates below expectation; green above. The Google Map is available at http://www.leydesdorff.net/nano2011/nano2011.htm.

Figure 6 shows that many cities other than Cambridge, UK, perform significantly above expectation, in Germany, for example. These, however, are often smaller towns with smaller universities, but at the aggregated level of countries Germany is ranked third and significantly above expectation, whereas the UK is ranked sixth and not different from expectation. The larger map informs us that papers from France and Mediterrenean countries on average have less impact in this field than expected.

## Specification of expectation

Because *I3*—and *mutatis mutandis PR6*—is based on a summation, the values for units of analysis such as countries or journals can be large. For example, the *I3*-value of *Nano Letters* in the set of 58 journals was 117,450, representing only 8% of the total *I3* of the set, namely 1,469,253. For the expected value of *I3*, this latter value was normalized using the share of publications in the set ( $n_i / \sum_i n_i$ = 1,507 / 31,644). Therefore, in this case: exp(*I3*) = 1,469,253 * (1,507 / 31,644) = 69,971. The differences between observed and expected values can thus be large; this so easily generates significance using the *z*-test that the test perhaps loses its discriminating value. The measure *I3* is very sensitive to differences in performance.[10]

An obvious alternative would be to test the percentage *I3* for each unit of analysis against the percentage of publications.[11] However, *I3* is dependent on the *N* of publications, and percentages (or permillages) may be small in the case of cities or institutes (< 0.1%). Using relative publication rates as expected values, for example, only *Nano Letters* and *Advances in Materials* were flagged as impacting significantly above expectation in

[10] The *z*-test is case-based. Bonferroni-Holm correction (Holm, 1979) can then be used for correcting family-wise accumulation of the probability of Type-I errors in multiple comparisons among cases by adjusting the $\alpha$-level (Rüdiger Mutz, *personal communication,* August 25, 2011). However, the accumulation in this case is not a consequence of multiple comparsions but of summation of *I3*-values within each case (e.g., city, journal).

[11] Leydesdorff & Bornmann (2011) used this expectation, but these authors used the standardized residuals of the chi-square for the test. The latter test is less conservative and less reliable than the *z*-test.

Table 1 ($p \leq 0.01$), but using this alternative option none of the smaller journals remain significantly above or below expectations. In Table 4, all significance indications for countries, cities, and institutes would disappear, and the Google map accordingly would become uniformative. This alternative is therefore unattractive.

A third option could be to compare the observed impact *per paper* with the expected one per paper. This specification prevents the accumulation of *I3* values and thus differences from expected values. The expected value of *I3*/paper on the basis of the set could then also be considered as a "world average." Instead of dividing the observed average by the expected one (as in the case of *MOCR/MECR* (Budapest), *NMCR* (Louvain), or the "old Leiden crown indicator"), the expected value could be used as a yardstick for the testing. However, one would expect the results to suffer from the same problems as the previous use of citations per publications, namely, that the *N* of papers is used in the denominator so that larger units of analysis are relatively disadvantaged and smaller units foregrounded.

One can draw the same map as Figure 6, but using this indicator (not shown here, but available at http://www.leydesdorff.net/nano2011/nano2011b.htm). Indeed, Cambridge UK (which is sized as before; $N = 228$) is no longer flagged as impacting significantly above expectation. Smaller centers such as Ludwigshaven in Germany ($N = 9$) and Maastricht in the Netherlands ($N = 8$) are now flagged as performing significantly above expectation. (A minimum of $N$ =5 was used for the test for reasons of reliability.)

Nevertheless, this map provides another view to the same data, and I decided to include it in the routine at the Internet. In addition to ztest.txt, a file ri3r.txt is generated by i3cit2.exe (and i3inst2.exe, respectively) which enables the user to draw this map. The z-score is also saved in ucities.dbf with the fieldname "zaverage."[12] Unlike previous indicators (derived from the Relative Citation Rate *RCR* = *MOCR*/*MECR*; Schubert & Braun, 1986), this indicator is not based on a division of means, but tests citations in terms of the percentile scores of the cited publications against a "world average" as the expectation. Let me call this indicator the *Relative I3-Rate* (*RI3R*) in honour of the *RCR* which has served as an indicator for 25 years.

In summary , one can distinguish between testing integrated (*I3*) and average (*I3/n)* citation impact. Integrated impact is more easily significantly different from expectation than average impact. Furthermore, the latter is also determined by the *n* in the denominator.

Let me finally note that one could further inform the expectation depending on the research question (Cassidy Sugimoto, *personal communication*, August 16, 2011). For example, one could assume that the chances of being cited are not equal for established authors and newcomers, and thus weigh publications according to the specification of this expectation. The statistical tests remain the same, but the values of the expectations are then different.

[12] Similarly, the file zperc.txt contains the overlay map when ‰*I3* is tested against ‰ of publications. The score is stored in the field "zperc" within ucities.dbf.

## Conclusions and discussion

The integrated impact indicator *I3* provides us with a versatile instrument which (*i*) can be applied to institutional units of analysis and journals (or other document sets); (*ii*) takes both publication and citation rates into account like the *h*-index; and (*iii*) enables us to use non-parametric statistics on skewed distributions as against using averages. The values of *I3* in terms of percentiles can (*iv*) be summed in evaluative schemes such as the six ranks used by the NSF (and otherwise; e.g. quartiles).

The first requirement is a careful decision about the reference set. This set can be based on a sophisticated search string in the Web of Science or on a delineated journal set. One can vary the citation windows and use fractional instead of integer counting. The measure is formalized much more abstractly, namely by reducing any set to one hundred percentiles and thus making unequal distributions comparable without giving up the notion of quality either by abandoning the tales of the distribution (as in the case of the *h*-index) or by using "total cites" instead of more sophisticated statistics.

Let me emphasize that the new measure does not prevent one from taking an average citation over publication rate as another measure. However, some of the issues in the debate over impact indicators that has raged for the past two years can be solved by using non-parametric statistics on which this new measure is based. I would particularly recommend its use combined with fractional citation counting as proposed by Leydesdorff & Opthof (2010) and hitherto applied in a limited number of studies (Leydesdorff & Bornmann, 2011a; Leydesdorff & Shin, 2010; Prathap, 2011; Zhou & Leydesdorff, 2011; Rafols *et al.*, in press; cf. Radicchi & Castellano, 2012). Fractional counting would correct for disciplinary structures (and biases) in the potentially interdisciplinary reference sets. In the case of nanoscience and nanotechnology, I have shown how one can identify 15 journals (out of 58) that are specific for condensed-matter and chemistry-based nanotechnology.

These analytical advantages can be combined with practicalities such as the straightforward option to show the results of an evaluation at the city or institute level by using Google Maps. As noted, fractional counting (in terms of authors and addresses) can be expected to further refine these maps. The software is available at http://www.leydesdorff.net/software/i3. This software routinely uses the *I3* based on 100 percentiles as well as the six percentile classes distinguished above.

**Normative implications**

The use of citation and publication analysis for evaluative purposes is a normative choice. Many good reasons can be given why one should not rely too much on statistics when taking decisions, particularly when the set is small, as in the case of individual hiring of promotion decisions. Peer review, however, also has constraints (e.g., Bornmann *et al.*, 2010), and in the case of large sets the reading of actual papers may be too time-consuming. Bibliometric indicators can then serve as proxies provided that error can be specified and the illusion of clarity conveyed by quantification can be avoided.

In Leydesdorff *et al.* (2011b, at pp. 1371f.), we specified a set of criteria for citation indicators, such as:

- Citation-based indicators should accommodate various normative schemes such as the six categories of the NSF;
- Citation-based indicators should be applicable to different sets of reference such as fields of science, nations, institutions, etc.;
- The indicator should allow productivity to be taken into account. One should, for example, be able to compare two papers in the 39$^{th}$ percentile with a single one in the 78$^{th}$ percentile (perhaps after weighing in an evaluation scheme);
- The indicator should result in a rather straightforward measure such as a percentage of maximum performance;
- Error estimation and statistics should enable the analyst to specify the significance of individual contributions and differences in rankings.

By developing the apparatus of *I3* and showing its comparability with the NSF scheme of six ranks, its decomposability in terms of percentages of contributions, and its relative straightforwardness in being based on percentiles in a distribution that allows for the use of non-parametric statistics, Leydesdorff & Bornmann (2011b) have shown how one can meet these criteria. This study has elaborated the technique for a highly policy-relevant field of science, notably, “nanoscience & nanotechnology,” and integrated *I3* with the geographic evaluation of excellence using Google Maps (Bornmann & Leydesdorff, 2011). The instrument is ready to be used.

**Acknowledgement**

I thank Lutz Bornmann for continuous discussion of these issues. I also wish to thank Cassidy Sugimoto, Ismael Rafols, and an anonymous referee for comments on a previous version of this paper. I am grateful to Thomson-Reuters for access to the data.